\documentclass[runningheads]{llncs}
\usepackage{appendix}
\usepackage[T1]{fontenc}
\usepackage{pythonhighlight}
\usepackage{amsmath,amssymb}
\usepackage{graphicx}
\usepackage{xcolor}
\usepackage{enumitem}
\usepackage{url}

\usepackage{amsmath}

\newif\ifComments

\Commentstrue

\begin{document}

\title{Abductive Vibe Coding (Extended Abstract)}

\author{Logan Murphy
\and
Aren A. Babikian
\and
Marsha Chechik
}
\authorrunning{L. Murphy et al.}

\institute{University of Toronto, Toronto, Canada}
\maketitle      
\begin{abstract}

When software artifacts are generated by AI models (``vibe coding''), human engineers assume responsibility for validating them. Ideally, this validation would be done through the creation of a formal proof of correctness. However, this is infeasible for many real-world vibe coding scenarios, especially when  requirements for the AI-generated artifacts resist formalization. This extended abstract describes ongoing work towards the extraction of analyzable, semi-formal \emph{rationales} for the adequacy of vibe-coded artifacts. Rather than deciding correctness directly, our framework produces a set of conditions under which the generated code can be considered adequate. We describe current efforts towards implementing our framework and anticipated research opportunities.
\end{abstract}

\subsubsection*{Introduction.}
One of the central impediments to the scalability of programming tools based on generative AI models (``coding agents'') is the need to \emph{trust} the artifacts they produce~\cite{lyu2025automatic}. One way of increasing trust is to request, alongside the code itself, an \emph{explanation} of why the code provides an adequate solution to the given problem. This explanation could be created in any number of ways; let us consider two extremes. In what we call the \emph{naive} approach, the explanation is given in unstructured natural language. Alternatively, in what we call the \emph{deductive} approach, the explanation is given as a formal proof of correctness. 

\begin{figure}
    \centering
    \includegraphics[width=0.8\linewidth]{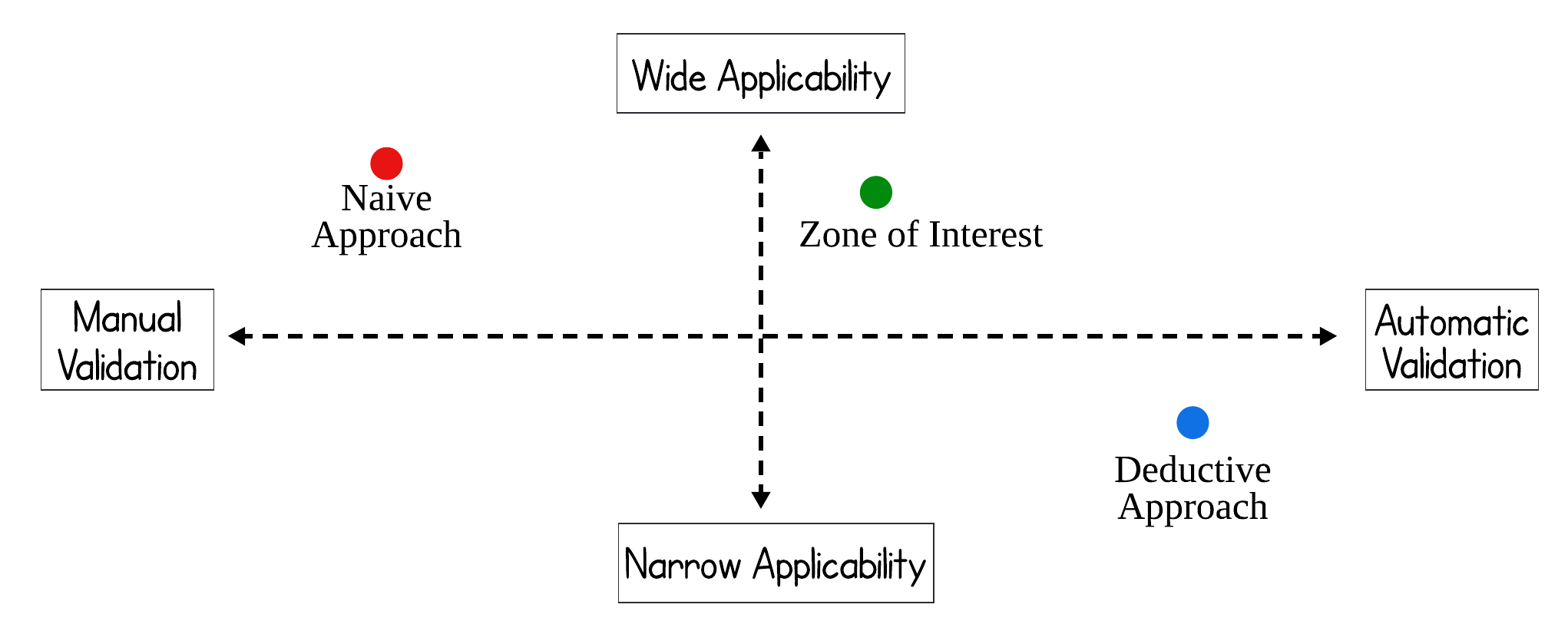}
    \caption{Comparing different approaches to creating explanations of adequacy for vibe-coded artifacts. Our proposal, \emph{abductive vibe coding}, aims to inhabit the ``Zone of Interest''. 
    }

        \label{fig:comparison}
\end{figure}

We can compare the strengths and weaknesses of these approaches across two dimensions, as shown in Fig.~\ref{fig:comparison}. The first dimension (the horizontal axis) considers how the explanation itself can be validated. Under the naive approach, the natural language explanation must be manually reviewed by the responsible engineer. This is of course error-prone, especially if the reviewer has an incentive to accept the explanation -- for instance, rejecting it could mean they will have to do some programming work. On the other hand, explanations created under the deductive approach can be automatically checked for correctness. The second dimension (the vertical axis) reflects the \emph{applicability} of the approach, i.e., the number of coding scenarios in which the approach could be applied. 
The naive approach has wide applicability (i.e., it can be done for virtually any use case), whereas the deductive approach requires both the semantics of the program \emph{and} its correctness specification to be formally interpretable. 

\subsubsection*{Motivating Example.}
Suppose that Alice is developing a software tool to support a financial firm's anti-money laundering (AML) procedures by automatically flagging accounts which exhibit certain risk factors. Alice wants a coding agent to draft a program which takes some information about a client account (transaction history, account age, existence of prior alerts, etc.) and, if appropriate, flags the account as suspicious. To this end, she writes a prompt enumerating, among other things, the following requirements for the code:
\begin{enumerate}[label=(\arabic*),topsep=0.3em]
    \item The code should return a tuple $\langle s, d, r\rangle$, where $s$  is in integer \emph{score} representing the overall risk identified for the account; $d \in \{flag, ~ok, ~review\}$ is a  \emph{decision} on whether the account should be flagged, ignored, or manually reviewed; and $r$ is a list of \emph{reasons} explaining how the value $s$ was computed for the account.
    \item The program logic involved in the computation of $s$ and $d$ should align with general AML priorities and best practices.
    \item The explanations provided in $r$ should be non-speculative and non-accusatory.
\end{enumerate}
Considering the above three points to form a ``specification'' for the code Alice wants generated, we emphasize that this specification resists a thorough formalization, especially with respect to points (2) and (3). Thus, if Alice obtains some code from an LLM which purports to meet this specification, it is unlikely that she would be able to validate it through a (traditional) proof of correctness. A sample vibe-coded solution to Alice's prompt is given in Appendix~\ref{app:code}.




\subsubsection*{Abductive Vibe Coding.}
We propose \emph{abductive vibe coding} as an alternative to both the {naive} and {deductive} approaches outlined above. The  ``abductive'' qualifier refers to the process of reasoning through \emph{hypothesis-generation} articulated by C.S. Peirce~\cite{niiniluoto1999defending}. The basic assumption upon which our approach builds is that the agent creating the explanation is \emph{entirely untrustworthy}, in the sense that it does not actually ``know'' any useful facts about the world, the program or the engineer's needs. The best the agent can do is \emph{hypothesize} facts which may or may not be true. Abductive vibe coding encourages the agent to explain the adequacy of some given code as a solution to the user's problem by generating \emph{hypotheses} which, if true, would imply the adequacy of the code. 

\begin{figure}[t]
    \centering

    \includegraphics[width=0.8\linewidth]{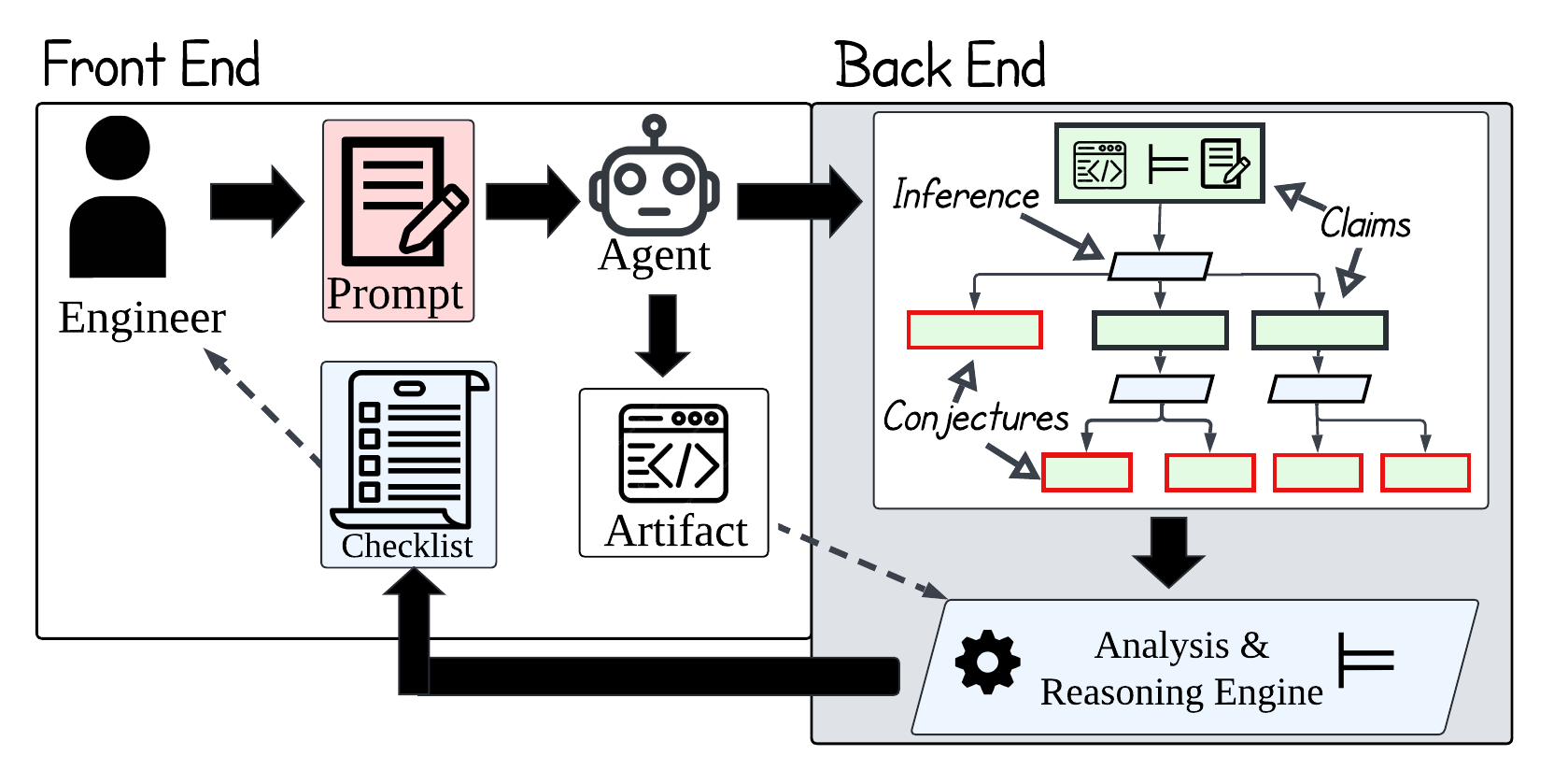}
    \caption{The ``abductive vibe coding'' workflow.}
    \label{fig:workflow}

    \end{figure}
In our implementation of abductive vibe coding, these explanations (which we refer to as \emph{rationales}) are given as \emph{structured}, \emph{hierarchical} and \emph{semiformal} arguments (shown on the right-hand-side of Fig.~\ref{fig:workflow}).
The design of our rationales is inspired by \emph{structured assurance cases} as expressed in languages such as Goal Structuring Notation (GSN)~\cite{kelly2004goal}. Structurally, such an argument takes the form of a rooted tree, whose vertices are primarily \emph{claims}, i.e., propositions hypothesized by the agent to be both true and useful in explaining the adequacy of the vibe-coded program. In our proposal, the claims in a rationale can be \emph{formal} (e.g., an interpretable sentence in a first-order theory), or \emph{semiformal} (e.g., containing some uninterpreted atomic propositions or predicates). The root node of the tree is a claim $C$ asserting that the vibe-coded artifact is an adequate solution to the given problem. This claim $C$ is then \emph{decomposed} into a set of subclaims $\{C_1,\ldots,C_n\}$, representing a purported \emph{inference} $\left(\bigwedge C_i\right) \implies C$.  This decomposition of claims into subclaims proceeds iteratively until a claim asserts something which ought to be supported by some observable \emph{evidence}. Such claims (called \emph{conjectures}) form the {leaves} of the tree. Thus, in creating a rationale, the agent iteratively decomposes claims by generating plausible hypotheses to use as subclaims, in such a manner that the resulting inferences should hold, until all remaining conjectures seem (to the agent) sufficiently precise to enable direct verification, e.g., via static analysis or by appealing to some external source of information.

Once the rationale has been generated, it is \emph{analyzed} in two respects: (i) each inference $\left(\bigwedge C_i\right) \implies C$ is checked for validity (e.g., using SMT solvers), and (ii) for each conjecture which can be soundly verified by some particular decision procedure (e.g., a program analysis tool), verification is attempted. Any of these tasks which do not succeed, as well as any conjectures for which no decision procedure is available, are returned to the user in the form of a \emph{checklist} of facts to be investigated offline. By construction, this checklist has the property that if all its items are \emph{certain} to hold, then we have a proof of the adequacy of the code with respect to the original specification. Of course, some items on the checklist may have a subjective element to them; in which case the user needs to determine their degree of \emph{confidence} or \emph{doubt} in their veracity. But because the items on the checklist are extracted directly from the rationale, the user can trace the  impact of any doubt onto the rest of the argument.

\subsubsection{Example (Continued).}
Continuing with the example from Alice and the AML program, suppose we ask the agent to produce a rationale for its adequacy as outlined above. A representative example is included in Appendix~\ref{app:rationale}, of which we only highlight some aspects. First, beginning with the root claim $C_R$ asserting the adequacy of the code, the rationale begins by proposing a ``shallow'' formalization of the specification, in which requirement (1) is formalized (as it only considers the structure of the program output), and requirements (2) and (3) are expressed using uninterpreted predicates; this formalization is used to define a single subclaim $C_0$, introducing the inference $C_0 \implies C_R$. Since $C_R$ is uninterpreted, this inference will be included in the checklist, i.e., Alice will need to ensure that the formalization adequately captures her intention for the specification. The claim $C_0$ is itself decomposed into three subclaims $\{C_1,C_2,C_3\}$, each of which asserts one of the three (semi)-formal requirements. Since $C_0$ is merely a conjunction of these three requirements, the resulting inference is provably valid and need not be reviewed by Alice. 

Claim $C_1$ describes the structure of the program output, which (in this case) is straightforward to statically verify from the source code. Claim $C_2$, which is concerned with the alignment of the program logic to AML practices, is itself decomposed into four subclaims, each of which represents an aspect of the program logic which the agent \emph{hypothesized} reflects alignment with AML practices. The inference resulting from this decomposition, which Alice will need to review, effectively asserts that these aspects of the logic are sufficient to demonstrate alignment with AML practices. At the discretion of the agent, these subclaims are either left as conjectures to be verified (e.g., ``The mapping of risk factors to risk tiers is informed by AML best practices and guidelines'') or themselves decomposed into additional subclaims.
Claim $C_3$, which asserts that the textual explanations produced by the program are non-speculative and non-accusatory, is decomposed into two subclaims: one of which identifies all string phrases that can be included in the program output, and one of which asserts that none of these concrete strings contains speculative or accusatory language. This decomposition is logically valid, and the completeness of the extracted strings relative to the program source code is statically verifiable, so Alice is left only to review each of the concrete strings for appropriateness.


\subsubsection{Towards Implementation.}
In order to implement our proposal, we need (i) a suitable format for encoding the rationales generated by agents, (ii) support for automatically analyzing these rationales, and (iii) effective strategies for prompting agents to produce useful rationales. Towards (i) and (ii), we are currently developing support for embedding rationales in the proof assistant Lean, building on our prior work on guided formalization of structured assurance cases~\cite{viger2023foremost}. Lean was chosen for its extensive metaprogramming facilities, which provides us flexibility in implementing both native and external analyses for rationales. With respect to (iii), we are simultaneously investigating various strategies and techniques for facilitating the creation of rigorous rationales, e.g., by developing partial  \emph{templates} for rationales which can be formalized \emph{a priori}.

\subsubsection{Research Opportunities.}
We believe that abductive vibe coding, as a general strategy, introduces several interesting research questions to be validated, and opportunities for extensions in various directions. There are various metrics by which the framework may be evaluated, e.g., its usability by engineers without formal methods knowledge, its effectiveness in increasing trust in AI-generated artifacts, the applicability of the framework to different vibe coding scenarios, and the costs of applying the framework (in terms of tokens or human involvement). We anticipate a wide range of possible strategies for improving the effectiveness of the framework across these and other dimensions, introducing several research opportunities. While the version of abductive vibe coding presented here takes place over a single iteration (artifact  $\to$ rationale $\to$ checklist), we believe that the benefits of the framework will be increased by extending it to an \emph{iterative} process. For instance, following a first draft of a rationale, an engineer may wish to prompt the model to reject some conjectures, or re-articulate or refine others. Alternatively, some of the conjectures generated by the agent may provide insight about how the engineer should refine their overall specification. Developing effective strategies for steering or fine-tuning the development of the rationale with iterative user feedback a very interesting and potentially significant research opportunity.
Finally, while our current proposal is primarily concerned with validating AI-generated code, we believe the approach could be generalized to support the validation of other engineering artifacts, such as specifications, models, tests scenarios, etc.


\subsubsection{Related Work.}

There have been several proposals for methods of validating AI-generated artifacts against informal specifications as a means to establish trust in AI systems. Proposed strategies include attempting to directly extract explanations from the the generated artifact~\cite{ruanSpecRoverCodeIntent2025}, 
performing semantic analysis on a given natural language explanation~\cite{nguyentungAutomatedTrustworthinessOracle2025},  
and autoformalizing user specifications to enable formally grounded evaluation~\cite{chenboqi2025constring,chenkua2023automated}. Our proposal for abductive vibe coding combines analysis and generation of informal specifications, automated formalization of specifications, and formal reasoning and analysis.

Human involvement in the verification and validation of AI-generated artifacts is an emerging area of investigation. Shankar et al.~\cite{shankarWhoValidatesValidators2024} incorporate human feedback on partial natural-language explanations, emphasizing human preferences rather than formal structure.
In a formal setting ~\cite{mitchellPositionVibeCoding2025}  integrates human feedback on formal reasoning alongside vibe coding workflows. Two distinguishing characteristics of our proposal are the focus on the logical structure of the rationale and the support for both formal and informal assertions about the artifact under evaluation.

\subsubsection{Conclusion.}
In this extended abstract, we introduced \emph{abductive vibe coding} as a means to develop rigorous and partially analyzable justifications for the adequacy of AI-generated software artifacts. Our proposal uses structured argumentation, supports a combination of formal and informal semantics, and encourages coding agents to generate hypotheses which can be independently verified. We believe that abductive vibe coding presents a promising path to improving the trustworthiness of AI-generated artifacts, especially for situations in which formal verification is not feasible. Moreover, we believe this proposal introduces several interesting opportunities for further research and development.

\bibliographystyle{splncs04}
\bibliography{refs,refs-aren}

@inproceedings{chenkua2023automated,
  title={Automated domain modeling with large language models: A comparative study},
  author={Chen, Kua and Yang, Yujing and Chen, Boqi and L{\'o}pez, Jos{\'e} Antonio Hern{\'a}ndez and Mussbacher, Gunter and Varr{\'o}, D{\'a}niel},
  booktitle={2023 ACM/IEEE 26th International Conference on Model Driven Engineering Languages and Systems (MODELS)},
  pages={162--172},
  year={2023},
  organization={IEEE}
}

@inproceedings{chenboqi2025constring,
  title = {{{LLM-based Satisfiability Checking}} of {{String Requirements}} by {{Consistent Data}} and {{Checker Generation}}},
  booktitle = {2025 {{IEEE}} 33rd {{International Requirements Engineering Conference}} ({{RE}})},
  author = {Chen, Boqi and Babikian, Aren A. and Feng, Shuzhao and Varr{\'o}, D{\'a}niel and Mussbacher, Gunter},
  year = 2025,
  pages = {231--243},
}

@inproceedings{mitchellPositionVibeCoding2025,
  title = {Position: {{Vibe Coding Needs Vibe Reasoning}}: {{Improving Vibe Coding}} with {{Formal Verification}}},
  shorttitle = {Position},
  booktitle = {Proceedings of the 1st {{ACM SIGPLAN International Workshop}} on {{Language Models}} and {{Programming Languages}}},
  author = {Mitchell, Jacqueline and Shaaban, Yasser},
  year = 2025,
  series = {{{LMPL}} '25},
  pages = {84--90},
  publisher = {Association for Computing Machinery},
}

@inproceedings{shankarWhoValidatesValidators2024,
  title = {Who {{Validates}} the {{Validators}}? {{Aligning LLM-Assisted Evaluation}} of {{LLM Outputs}} with {{Human Preferences}}},
  shorttitle = {Who {{Validates}} the {{Validators}}?},
  booktitle = {Proceedings of the 37th {{Annual ACM Symposium}} on {{User Interface Software}} and {{Technology}}},
  author = {Shankar, Shreya and {Zamfirescu-Pereira}, J.D. and Hartmann, Bjoern and Parameswaran, Aditya and Arawjo, Ian},
  year = 2024,
  series = {{{UIST}} '24},
  pages = {1--14},
  publisher = {Association for Computing Machinery},
}

@article{nguyentungAutomatedTrustworthinessOracle2025,
  title = {Automated {{Trustworthiness Oracle Generation}} for {{Machine Learning Text Classifiers}}},
  author = {Nguyen Tung, Lam and Cho, Steven and Du, Xiaoning and Neelofar, Neelofar and Terragni, Valerio and Ruberto, Stefano and Aleti, Aldeida},
  year = 2025,
  journal = {Proc. ACM Softw. Eng.},
  volume = {2},
  number = {FSE},
  pages = {2382--2405},
}

@inproceedings{ruanSpecRoverCodeIntent2025,
  title = {{{SpecRover}}: {{Code Intent Extraction}} via {{LLMs}}},
  shorttitle = {{{SpecRover}}},
  booktitle = {Proceedings of the {{IEEE}}/{{ACM}} 47th {{International Conference}} on {{Software Engineering}}},
  author = {Ruan, Haifeng and Zhang, Yuntong and Roychoudhury, Abhik},
  year = 2025,
  series = {{{ICSE}} '25},
  pages = {963--974},
}

@article{niiniluoto1999defending,
  title={Defending abduction},
  author={Niiniluoto, Ilkka},
  journal={Philosophy of science},
  volume={66},
  number={S3},
  pages={S436--S451},
  year={1999},
  publisher={Cambridge University Press}
}

@article{viger2023foremost,
  title={The foremost approach to building valid model-based safety arguments},
  author={Viger, Torin and Murphy, Logan and Di Sandro, Alessio and Menghi, Claudio and Shahin, Ramy and Chechik, Marsha},
  journal={Software and Systems Modeling},
  volume={22},
  number={5},
  pages={1473--1494},
  year={2023},
  publisher={Springer}
}

@article{lyu2025automatic,
  title={Automatic programming: Large language models and beyond},
  author={Lyu, Michael R and Ray, Baishakhi and Roychoudhury, Abhik and Tan, Shin Hwei and Thongtanunam, Patanamon},
  journal={ACM Transactions on Software Engineering and Methodology},
  volume={34},
  number={5},
  pages={1--33},
  year={2025},
  publisher={ACM New York, NY}
}

@inproceedings{kelly2004goal,
  title={The goal structuring notation--a safety argument notation},
  author={Kelly, Tim and Weaver, Rob},
  booktitle={Proceedings of the dependable systems and networks 2004 workshop on assurance cases},
  volume={6},
  year={2004},
  organization={Citeseer Princeton, NJ}
}
\newpage
\appendix
\section{Sample Vibe-Coded Program}
\label{app:code}
\begin{python}
def assess_suspicious_activity(account, risk_factors):
    reasons = []
    score = 0.0

    LOW_RISK_WEIGHT = 1.0
    MID_RISK_WEIGHT = 3.0
    HIGH_RISK_WEIGHT = 6.0

    SUSPICIOUS_THRESHOLD = 8.0
    AMBIGUOUS_THRESHOLD = 4.0

    MAX_MITIGATION = 4.0  

    transactions = risk_factors.get("transactions")
    account_age_days = risk_factors.get("account_age_days")
    customer_profile = risk_factors.get("customer_profile")
    prior_alerts = risk_factors.get("prior_alerts")

    high_risk_jurisdictions = set(high_risk_countries())

    high_risk_txn_count = 0
    for txn in transactions:
        if txn.get("country") in high_risk_jurisdictions:
            high_risk_txn_count += 1

    if high_risk_txn_count > 0:
        score += HIGH_RISK_WEIGHT
        reasons.append("Transactions involving higher-risk 
              jurisdictions were observed ")

    large_txn_count = sum(1 for txn in transactions 
                          if txn.get("amount", 0) >= 100000)

    if large_txn_count >= 2:
        score += MID_RISK_WEIGHT
        reasons.append("Multiple high-value transactions were 
        recorded within the review period.")

    if account_age_days < 90:
        score += MID_RISK_WEIGHT
        reasons.append("Account is relatively new, which may 
        limit historical context for activity patterns.")

    if prior_alerts > 0:
        score += MID_RISK_WEIGHT
        reasons.append("Prior monitoring alerts exist for 
        this account.")

    if len(transactions) > 20:
        score += LOW_RISK_WEIGHT
        reasons.append("Transaction volume is elevated 
        compared to typical baseline activity.")

    if "cash-intensive" in customer_profile:
        score += LOW_RISK_WEIGHT
        reasons.append("Features of customer profile is 
        associated with moderately elevated AML monitoring 
        sensitivity.")

    mitigation_score = mitigation_kb(account, risk_factors)
    mitigation_offset = min(mitigation_score, MAX_MITIGATION)
    score -= mitigation_offset

    if mitigation_offset > 0:
        reasons.append("Documented contextual factors may 
        explain some observed activity.")

    score = max(score, 0.0)

    if score >= SUSPICIOUS_THRESHOLD:
        decision = "flag"
    elif score >= AMBIGUOUS_THRESHOLD:
        decision = "review"
    else:
        decision = "ok"

    return {
        "score": round(score, 2),
        "decision": decision,
        "reasons": reasons
    }
\end{python}

\newpage 
\section{Sample Rationale}
\label{app:rationale}
Let $A$ be the set of valid inputs to the program. Let $f$ represent the program about which the argument is expressed. We outline here the logical structure of a sample rationale for our running example. {The  rationale described here, as well as the program it describes, were generated using the free version of OpenAI GPT 5.2, through some manual iteration.}

\vspace{2mm}
\noindent
\textbf{Root Claim $C_R$:} The program satisfies the given specification. 

\noindent 
\textbf{Decomposed by}: $\{C_0\}$

\noindent
\textbf{Note}: This decomposition must be validated by the user.

\vspace{3mm}
\noindent 
\textbf{Claim $C_0$:} $\forall x \in A, R_1(f(x)) \wedge R_2(x,f(x)) \wedge R_3(f(x))$
where 
\begin{itemize}[itemsep=0.5em]
    \item $R_1(v)$ asserts $v = \langle s, d,r\rangle$ for some $s \in \mathbb{R}$, $d \in \{flag,~ok,~review\}$ and $r \in List(String)$.
    \item $R_2(x,v)$ asserts that the score and decision described by output $v$ reflect general AML guidelines and best practices as they would be applied to the input $x$.
    \item $R_3(f(x))$ asserts that any explanations included in the output are non-accusatory and non-speculative. Formally, $R_3(v) \equiv \forall s \in reasons(v), \neg Accusatory(s) \wedge \neg Speculative(s)$, where $Accusatory(s)$ and $Speculative(s)$ are left formally uninterpreted.
\end{itemize}
\textbf{Decomposed by} $\{C_1,C_2,C_3\}$ 

\noindent
\textbf{Note}: This decomposition is logically valid.

\begin{itemize}[itemsep=1em]
    \item[] \textbf{Conjecture $C_1$}: $\forall x \in A, R_1(f(x))$ 

    \textbf{Note:} This conjecture can be verified through static analysis of the program.

    \item[] \textbf{Claim $C_2$}: $\forall x \in A, R_2(x,f(x))$

\textbf{Decomposed by}: $\{C_4,C_5,C_6,C_7\}$

\noindent
\textbf{Note}: This decomposition must be validated by the user.

\begin{itemize}[topsep=0.8em]
    \item[] \textbf{Conjecture $C_4$}:  Each risk factor is mapped to exactly one of three risk tiers $\{low, mid, high\}$ 
    
    \item[] \textbf{Conjecture $C_5$}:  The mapping of risk factors to risk tiers is informed by AML best practices and guidelines 
    
    \item[] \textbf{Conjecture $C_6$}:  $\forall a,b \in A. ~score(f(a)) \leq score(f(b)) \to decision(f(a)) \leq decision(f(b))$ 

    \textbf{Note}: This asserts that the risk decision is monotonic with respect to the score, under ordering $ok <review < flag$. This can be verified by static analysis.

    \item[] \textbf{Claim $C_7$}: Weights associated to risk tiers and thresholds for decisions are appropriate.
    
    \textbf{Decomposed by}: $\{C_8, C_{9}, C_{10}\}$

    \textbf{Note}: This decomposition must be validated by the user. Intuitively, this inference asserts that $C_8$, $C_{9}$ and $C_{10}$ form a sufficient set of requirements for the weights associated to the risk tiers and threshold for decisions are appropriate.

    \begin{itemize}[topsep=0.8em]
        \item[] \textbf{Conjecture} $C_8$: Mid-tier risk factors are given three times the significance of a low-tier risk factor, and high-risk tier factors are given twice the significance of a mid-tier risk factor. 

        \item[] \textbf{Claim} $C_{9}:$ The ratio of risk factor weight ($low \mapsto 1.0, mid \mapsto 3.0, high\mapsto 6.0$), decision thresholds $(review \mapsto 4.0, flag \mapsto 8.0)$, and maximum mitigation factor ($-4.0$) is appropriate and aligned with best AML practices.
        
        \item[] \textbf{Claim} $C_{10}:$ The knowledge base (``${\tt mitigation\_{kb}}$'') provides an appropriate mitigation score for the given account and risk factors.
        
    \end{itemize}
\end{itemize}
 \item[] \textbf{Claim $C_3$}: $\forall x \in A, R_3(f(x))$

Equivalently: $\forall x \in A, \forall s \in reasons(f(x)), \neg Accusatory(s) \wedge \neg Speculative(s)$

\textbf{Decomposed by}: $\{C_{11}, C_{12}\}$

\textbf{Note}: This decomposition is logically valid.

\begin{itemize}[topsep=0.8em]
    \item[] \textbf{Conjecture $C_{11}$}: $\forall x \in A, \forall s, s \in reasons(f(x)) \implies s \in \{s_1,\ldots,s_7\}$, where 
    \begin{itemize}
        \item[] $s_1 = $ ``Transactions involving higher-risk jurisdictions were observed'' 
        \item[] $s_2 = $ ``Multiple high-value transactions were recorded within the review period''
        \item[] ... 
        \item[] $s_7 = $ ``Documented contextual factors may explain some observed activity''
    \end{itemize}
    \vspace{2mm}
    \textbf{Note:} This conjecture can be statically verified.
    \item[] \textbf{Conjecture} $C_{12}$: $\forall s \in \{s_1,...,s_7\}, \neg Accusatory(s) \wedge \neg Speculative(s)$
\end{itemize}


\end{itemize}
\vspace{4mm}
\noindent 

\vspace{3mm}
\noindent

\end{document}